\documentclass[12pt]{article}
\usepackage{graphicx}
\newcommand{\bb}{\begin{equation}}
\newcommand{\ee}{\end{equation}}
\newcommand{\ba}{\begin{eqnarray}}
\newcommand{\ea}{\end{eqnarray}}

\begin{document}

\title{{\bf Discrete Orbit Effect\\
Lengthens Merger Times for\\
Inspiraling Binary Black Holes}}

\author{
Don N. Page
\thanks{Internet address:
profdonpage@gmail.com}
\\
Department of Physics\\
4-183 CCIS\\
University of Alberta\\
Edmonton, Alberta T6G 2E1\\
Canada
}

\date{2024 June 7}

\maketitle
\large
\begin{abstract}
\baselineskip 20 pt

The inspiral merger time for two black holes captured into a nonrelativistic bound orbit by gravitational radiation emission has been often calculated by a formula of Peters that assumes the adiabatic approximation that the changes per orbit are small.  However, initially this is not true for the semimajor axis and period of most of the initially highly eccentric orbits, which change significantly during closest approach and much less elsewhere along the orbit.  This effect can make the merger time much longer (using other formulas from Peters that do not assume the adiabatic approximation) than that calculated by the adiabatic formula of Peters.

\end{abstract}

\normalsize

\baselineskip 22 pt

\newpage

\section{Introduction}

The motivation for studying the inspiraling of two black holes has increased greatly with the enormous success that LIGO and Virgo have had in detecting the gravitational waves from astrophysical black holes that coalesce to form a larger black hole
\cite{
LIGOScientific:2016aoc,
LIGOScientific:2016emj,
LIGOScientific:2016vbw,
LIGOScientific:2016vlm,
LIGOScientific:2016kwr,
LIGOScientific:2016vpg,
LIGOScientific:2016sjg,
LIGOScientific:2016dsl,
LIGOScientific:2016wyt,
LIGOScientific:2017bnn,
LIGOScientific:2017ycc,
LIGOScientific:2017vox,
LIGOScientific:2018mvr,
LIGOScientific:2018jsj,
Venumadhav:2019lyq,
LIGOScientific:2019lzm,
LIGOScientific:2020aai,
LIGOScientific:2020stg,
LIGOScientific:2020zkf,
LIGOScientific:2020iuh,
LIGOScientific:2020ufj,
LIGOScientific:2020ibl,
LIGOScientific:2021usb,
LIGOScientific:2021djp}.

In order to learn about the possible origin and evolution of the coalescing  black holes, many scenarios have been investigated, far too many for me to cite here, though a sample relevant to my considerations is
\cite{Peters:1963ux,Peters:1964qza,Peters:1964zz,Hansen:1972jt,Turner1977,Quinlan:1989,Lee1993,Cutler:1994pb,Sigurdsson:1994ju,Mouri2002,Jaffe:2002rt,OLeary:2005vqo,Pretorius:2007nq,OLeary:2008myb,Wen:2011xc,Loutrel:2014vja,Loutrel:2016cdw,Clesse:2016ajp,Kovetz:2016kpi,Loutrel:2017fgu,Raidal:2017mfl,Ali-Haimoud:2017rtz,Gondan:2017wzd,Loutrelthesis2018,Sasaki:2018dmp,DOrazio:2018jnv,Barack:2018yly,Rodriguez:2018pss,Raidal:2018bbj,Samsing:2019dtb,Vaskonen:2019jpv,Loutrel:2019kky,Korol:2019jud,Jedamzik:2020ypm,Kritos:2020fjw,Gerosa2021,Tucker:2021mvo,Kritos:2022ggc,Kocsis2022}.
Instead, I wish to focus on scenarios in which two isolated black holes (i.e., not originally a bound binary pair) become bound through the emission of gravitational radiation and then inspiral primarily through the process of emitting more gravitational radiation.  That is, I shall ignore the effects of any third-party masses on the formation and evolution of the bound pair, so that after capture into a bound orbit, the size of the orbit is assumed to be much smaller than the distance to any other objects of similar or greater mass, which does put a strong restriction on the astrophysical sites where my idealizations give an adequate approximation.

To set the notation, I shall adopt the center-of-momentum frame and start with two initially unbound black holes of masses $M_1$ and $M_2$ far away from each other\footnote{Note below that even in the formulas for the Keplerian orbits, $v$ or $\beta$ without a subscript is always the initial relative velocity or its dimensionless form when the black holes were unbound and {\it not} any velocity during an orbit; I use instead $v_p$ for the velocity, and $\beta_p = v_p/c$ as its dimensionless form, at periapsis, the point of smallest separation distance, $r_p$, during an orbit.}, 
\bb
\mathrm{total\ mass\ } \equiv M \equiv M_1 + M_2, 
\label{M}
\ee 
\bb
\mathrm{reduced\ mass\ } \equiv \mu \equiv \frac{M_1 M_2}{M_1+M_2} \leq \frac{1}{4}M,
\label{mu}
\ee
\bb
\mathrm{symmetric\ mass\ ratio\ } \equiv \eta \equiv \frac{\mu}{M} \equiv \frac{M_1M_2}{(M_1+M_2)^2} \leq \frac{1}{4}, 
\label{eta}
\ee
\bb
\mathrm{initial\ relative\ velocity\ } \equiv v \ll \eta^{1/2} c,
\label{v}
\ee
\bb
\mathrm{initial\ dimensionless\ relative\ velocity\ } \equiv \beta \equiv \frac{v}{c} \ll \eta^{1/2}, 
\label{beta}
\ee
\bb
\mathrm{initial\ nonrelativistic\ energy\ } \equiv E_0 = \frac{1}{2}\mu v^2 = \frac{1}{2}Mc^2\eta \beta^2 \ll \eta\mu c^2,
\label{E0}
\ee
\bb
\mathrm{relative\ impact\ parameter\ } \equiv b \gg \frac{4GM}{cv},
\label{b}
\ee
\bb
\mathrm{initial\ angular\ momentum\ }L_0 = \mu b v = M c b \eta \beta,
\label{L}
\ee
and
\bb
\mathrm{convenient\ timescale\ } \equiv T_0 \equiv \frac{2\pi GM}{v^3} 
= (0.264\ \mathrm{year})\left(\frac{M}{10M_\odot}\right)
\left(\frac{100\ \mathrm{km/s}}{v}\right)^3.
\label{T0}
\ee

I assume that the impact parameter is below the critical value $b_c$ (given by Eq.\ (\ref{bc}) below), so that the two black holes are captured into a bound orbit by emitting gravitational radiation, in the approximation of otherwise empty spacetime, and I assume that the bound orbit is initially nonrelativistic, which requires the strong inequality given above that $b \gg 4GM/(cv)$.  For $4GM/(cv) \ll b < b_c$, it turns out that I need to assume the other further strong inequalities given above that $\beta \ll \sqrt{\eta}$ or initial relative velocity $v \ll c \sqrt{\eta} \equiv c\sqrt{M_1M_2}/(M_1+M_2)$.

I further assume that the two black holes are approaching each other with a uniform probability distribution in the transverse plane (probability per small range of impact parameter proportional to the impact parameter).  There is then a critical impact parameter $b_c$, given by Eq.\ (\ref{bc}) below, proportional to the Schwarzschild radius of the total mass $M = M_1 + M_2$, multiplied by the 7th root of the symmetric mass ratio $\eta$, and divided by the 9/7 power of the dimensionless initial velocity $\beta$, such that for impact parameters $b$ larger than $b_c$, the two black holes scatter without becoming bound, but for $b < b_c$, the black holes emit enough radiation in their initial close encounter that they become bound.  Conditionalizing on $b < b_c$, the cumulative probability for an impact parameter smaller than $b$ is then 
\bb
P = \left(\frac{b}{b_c}\right)^2,
\label{P}
\ee 
the fraction $(b/b_c)^2$ of the total area inside the capture disk of radius $b_c$ that also has the impact parameter vector of magnitude smaller than $b$.  This cumulative conditional probability $P$ then serves as a convenient dimensionless parameter characterizing the impact parameter $b$ when it is less than the critical impact parameter $b_c$ so that the two black holes become bound by the loss of energy to gravitational radiation.  

As a result, the coordinate-independent properties of the $n$th orbit (taken to begin at the $n$th periapsis and end at the $(n+1)$st one) are determined by the total mass $M$, the symmetric mass ratio $\eta$, the initial dimensionless velocity $\beta = v/c$, the dimensionless cumulative conditional probability $P = (b/b_c)^2$, the number $n$ of the orbit, and whatever combination of $G$ and $c$ that are needed to give the dimensions of the dimensional quantities.  One such quantity is
\bb
\mathrm{the\ {\it{n}}th\ orbit\ time\ period\ } \equiv \tau_n = 2\pi\sqrt{\frac{a_n^3}{GM}},
\label{taun}
\ee
with $a_n$ the semimajor axis of the $n$th nonrelativistic Keplerian orbit.  Then
\bb
\mathrm{total\ merger\ inspiral\ time\ } \equiv T \equiv \sum_{n} \tau_n
\label{T}
\ee
where the sum is taken over all the orbits until coalescence.  (For impact parameter $b \gg 4GM/(cv)$ as I assume, the number of nonrelativistic orbits will be large, but with decreasing periods $\tau_n$, so an approximate formula for this decrease of the periods can be summed as if it were an infinite series, since the error of the terms with large $n$ will be small in comparison with the contribution from the first several $n$.  But the absolute error will presumably go as $(2GM/c^3)/\beta^s$ with some positive exponent $s$ that is smaller than the $s=3$ exponent that applies for the actual timescale that will be of the order of $T_0 = (2\pi GM/c^3)/\beta^3$ multiplied by some dimensionless function of $P$.) 

Calculations of the capture and subsequent inspiraling of the black holes usually start with the formulas of Philip Carl Peters and of Peters and Jon Mathews \cite{Peters:1963ux,Peters:1964qza,Peters:1964zz}, who show that for a nonrelativistic Keplerian orbit of semimajor axis $a$ and eccentricity $e$, the power radiated into gravitational waves, averaged over one orbit, is (Eq.\ (16) in \cite{Peters:1963ux}, the same copied on page 92 of \cite{Peters:1964qza}, and Eq.\ (5.4) of \cite{Peters:1964zz})
\ba
\left\langle -\frac{dE}{dt}\right\rangle &=&
\frac{32}{5}\frac{G^4}{c^5}\frac{M_1^2 M_2^2(M_1+M_2)}{a^5(1-e^2)^{7/2}}
\left(1 + \frac{73}{24}e^2 + \frac{37}{96}e^4\right) \nonumber \\
&=& \frac{32}{5}\frac{G^4}{c^5}\frac{M^5\eta^2}{a^5(1-e^2)^{7/2}}
\left(1 + \frac{73}{24}e^2 + \frac{37}{96}e^4\right).
\label{power}
\ea
Similarly, the angular momentum radiated into gravitational waves, averaged over one orbit, is (Eq.\ (5.32) on page 99 of \cite{Peters:1964qza}, and Eq.\ (5.5) of \cite{Peters:1964zz})
\ba
\left\langle -\frac{dL}{dt}\right\rangle &=&
\frac{32}{5}\frac{G^{7/2}}{c^5}\frac{M_1^2 M_2^2(M_1+M_2)^{1/2}}
{a^{7/2}(1-e^2)^2}
\left(1 + \frac{7}{8}e^2\right) \nonumber \\
&=& \frac{32}{5}\frac{G^{7/2}}{c^5}\frac{M^{9/2}\eta^2}{a^{7/2}(1-e^2)^2}
\left(1 + \frac{7}{8}e^2\right).
\label{torque}
\ea

Since a Keplerian elliptical orbit has energy $E = -GM_1 M_2/(2a) = -GM^2\eta/(2a)$, angular momentum $L = \sqrt{GM^3\eta^2 a(1-e^2)}$, and orbital period $\tau = 2\pi\sqrt{a^3/(GM)}$, the energy and angular momentum radiated during one orbit are
\ba
-\Delta E &=& \tau\left\langle -\frac{dE}{dt}\right\rangle 
 = \frac{64\pi}{5}\frac{G^{7/2}M^{9/2}\eta^2}{c^5 a^{7/2}(1-e^2)^{7/2}}
\left(1 + \frac{73}{24}e^2 + \frac{37}{96}e^4\right),
\label{DeltaE}\\
-\Delta L &=& \tau\left\langle -\frac{dL}{dt}\right\rangle
 = \frac{64\pi}{5}\frac{G^3 M^4 \eta^2}{c^5 a^2 (1-e^2)^2}
\left(1 + \frac{7}{8}e^2\right).
\label{DeltaL}
\ea
Then for simplicity defining
\bb
f \equiv 1-e^2 = \frac{-2EL^2}{G^2M^5\eta^3}
\label{f}
\ee
(the square of the ratio of the semiminor axis $a(1-e^2)^{1/2}$ to the the semimajor axis $a$ of the Keplerian elliptical orbit), one gets the dimensionless ratios
\bb
-\frac{\Delta E}{E_0} = \frac{340\pi}{3}\left(\frac{GM^2\eta}{cL}\right)^7 \beta^{-2}
\eta\left(1-\frac{366}{425}f+\frac{37}{425}f^2\right),
\label{DeltaE/E_0}
\ee
\bb
-\frac{\Delta L}{L} = 24\pi\left(\frac{GM^2\eta}{cL}\right)^5 \eta\left(1-\frac{7}{15}f\right),
\label{Deltal/l}
\ee
which are now in a form that can be applied even for the initially hyperbolic orbit.

For the two initially unbound black holes with positive initial nonrelativistic energy $E_0 = (1/2)\mu v^2$ (initially very far apart, with negligible potential energy) to become bound into a nonrelativistic Keplerian orbit with negative nonrelativistic energy $E_1 = -GM_1M_2/(2a_1) = -GM^2\eta/(2a_1)$ (including the Newtonian potential energy $-GM_1M_2/r = -GM^2\eta/r$), the value of $-\Delta E/E_0$ during the transition between the initial hyperbolic orbit and the first elliptical orbit (taken to begin at the initial periapsis) must be greater than 1, so the initial angular momentum $L_0 = \mu b v$ must have the impact parameter $b$ be less than a critical impact parameter $b_c$ (depending linearly on the Schwarzschild radius $2GM/c^2$ of the total mass with a coefficient that is a pure number multiplying powers of the symmetric mass ratio, $\eta$, and of the dimensionless relative initial velocity, $\beta = v/c$, to be given shortly). 

One can readily show that this requirement of $-\Delta E/E_0 > 1$ with $\beta \ll 1$ implies that the initial hyperbolic orbit must have its eccentricity $e$ just slightly larger than 1 (a necessary but not sufficient condition for capture when $v \ll c$ or $\beta \ll 1$), so the initial $f = 1 - e^2$ is just slightly negative.  Conversely, for the initial elliptical orbit after $E$ becomes negative to be nonrelativistic (periapsis distance much larger than the Schwarzschild radius corresponding to the total mass $M$), one needs the eccentricity $e$ right after binding to be just slightly smaller than 1, so for the initial orbits at least, $f$ is just slightly positive.  These facts allow one to drop the terms with nonzero $f$ in the expressions above for the initial orbit and easily to get from them a good approximation for the critical impact parameter,
\bb
b_c = \left(\frac{85\pi}{384}\right)^{1/7}\left(4\eta\right)^{1/7}\beta^{-9/7}\left(\frac{2GM}{c^2}\right) 
\approx 0.949\,429\left(4\eta\right)^{1/7}\beta^{-9/7}\left(\frac{2GM}{c^2}\right).
\label{bc}
\ee
Here I have used $4\eta = 4 M_1M_2/(M_1+M_2)^2$ with the factor of 4 so that its upper bound is 1, and I have included the 2 in $2GM/c^2$ so that it is the Schwarzschild radius of the total mass $M$, which coincidentally makes the leading numerical factor only about 5\% less than of unity.

After I independently derived Eq.\ (\ref{bc}) above (see \cite{Page:2024zpr} for more details) from the equations of Peters and Mathews \cite{Peters:1963ux,Peters:1964qza,Peters:1964zz}, I was informed of the 1972 paper of R.\ O.\ Hansen \cite{Hansen:1972jt}, which gave two complicated formulas relating the initial energy, initial eccentricity, and initial impact parameter for the critical cases (with fixed masses), but it did not use the fact that one can neglect $f = 1 - e^2$ to get the explicit formula above for the critical impact parameter in terms of the initial relative velocity $v$ (and hence initial energy $E_0 = (1/2)\mu v^2$, for fixed given masses).  It was also pointed out to me that the 1979 paper of Martin Walker and Clifford Will \cite{Walker:1979xk} gave an explicit formula for the maximum eccentricity of the initial unbound hyperbolic orbit that would lead to capture in terms of the semilatus rectum of this orbit, which can be readily transformed to give the formula above for the critical impact parameter in terms of the initial relative velocity.  However, I found that the formula in essentially the same form as that above was given by the 1989 paper of Gerald D.\ Quinlan and Stuart L.\ Shapiro \cite{Quinlan:1989}.  Hideaki Mouri and Yoshiaki Taniguchi \cite{Mouri2002} also got the same formula but from the starting point of calculations of gravitational radiation from parabolic orbits by Michael Turner \cite{Turner1977} rather than from Peters and Mathews \cite{Peters:1963ux,Peters:1964qza,Peters:1964zz}.   A later paper giving the formula (presumably independently, since not mentioning that Quinlan and Shapiro had given it, though they were cited for other reasons) is \cite{OLeary:2008myb}.

All of the papers I found
\cite{Lee1993,OLeary:2008myb,Clesse:2016ajp,Kovetz:2016kpi,Ali-Haimoud:2017rtz,Gondan:2017wzd,Sasaki:2018dmp,Rodriguez:2018pss,Raidal:2018bbj,Vaskonen:2019jpv,Korol:2019jud,Jedamzik:2020ypm,Kritos:2020fjw,Tucker:2021mvo,Kocsis2022}
using the formula for the critical impact parameter $b_c$, as a starting point for calculating the inspiral merger time, used Eq.\ (5.14) of \cite{Peters:1964zz},
or the unnumbered last equation of that paper by Peters that is the approximation for initial eccentricity $e_0$ near 1 (equivalent to my $f_1 \equiv 1-e_1^2 \ll 1$ for the first bound orbit),
\bb
T(a_0,e_0) \approx (768/425) T_c(a_0) (1-e_0^2)^{7/2},
\label{Peters-eccentric}
\ee
where Eq.\ (5.10) of that paper gave the decay time for a circular orbit as
\bb
T_c(a_0) = a_0^4/(4\beta_P),
\label{Peters-circular}
\ee
with the preceding unnumbered equation giving
\bb
\beta_P = \frac{64}{5}\frac{G^3 m_1m_2(m_1+m_2)}{c^5}
\label{Peters-beta}
\ee
(and with Peters' $m_1$ and $m_2$ being the same as my $M_1$ and $M_2$; here I have added the subscript $P$ to Peters' $\beta$, which has the dimension of volume times velocity, so that it would not be confused with my dimensionless $\beta$ for the initial velocity or $\beta_p$ for the dimensionless velocity at periapsis below).  

However, this Eq.\ (\ref{Peters-eccentric}) assumes an adiabatic approximation for the evolution of the semimajor axis $a$ and resulting orbit period that I shall call $\tau$,
\bb
\tau = 2\pi\sqrt{\frac{a^3}{GM}},
\label{tau}
\ee
namely, that the fractional change in $a$ and $\tau$ from one orbit to the next is small.  This approximation is not valid for black hole binaries that become bound by the emission of gravitational radiation (a situation not discussed by Peters, so he can hardly be faulted for not pointing out that his adiabatic approximation would not apply in such cases).  Instead, both the length (major axis) $2a$ and the period $\tau$ of several of the initial orbits after the black holes become bound change significantly from one orbit to the next, and Eq.\ (\ref{Peters-eccentric}) severely underestimates the total merger inspiral time, especially when the impact parameter $b$ is just slightly smaller than the critical impact parameter given by Eq.\ (\ref{bc}).

\section{Properties of the Orbits and Total Merger Time}

We have noted above that in principle the dimensionless properties of the $n$th orbit, in our idealized situation of two black holes coming in nonrelativistically from far away with no other perturbing masses around, are determined by the orbit number $n$, the symmetric mass ratio $\eta = M_1M_2/(M_1+M_2)^2$, the dimensionless initial relative velocity $\beta = v/c \ll \eta^{1/2}$, and the cumulative conditional probability $P = (b/b_c)^2 < 1$ (the square of the ratio of the actual impact parameter $b$ to the critical impact parameter $b_c$ for capture into a bound orbit by the emission of gravitational radiation).  However, to get a quantative analysis of the orbits, it is helpful also to define some other dimensionless constants and variables.  After completing this section I found that some, but certainly not all, of the first part of this section has similarities with the analysis of \cite{Gondan:2017wzd}, though that paper does not discuss the merger time.

First are two dimensionless constants that only depend on the symmetric mass ratio $\eta$ and on the dimensionless initial relative velocity $\beta$:
\bb
\beta_c \equiv \left(\frac{96}{85\pi}\beta^2\frac{1}{\eta}\right)^{1/7} = \frac{2GM}{c^2 \beta b_c} = \frac{2GM/(cv)}{b_c},
\label{beta_c}
\ee
\bb
\delta \equiv \frac{2\beta}{\beta_c} 
\equiv \left(\frac{340\pi}{3}\beta^5\eta\right)^{1/7} = \frac{b_c\beta^2}{GM/c^2} = \frac{b_c}{GM/v^2}
= \left(\frac{85\pi}{24}\eta\right)^{1/2} \beta_c^{5/2}.
\label{delta}
\ee

The constant $\beta_c$ is the dimensionless velocity $\beta_p = v_p/c$ at the periapsis of the first orbit in the limit that the impact parameter $b$ approaches the critical impact parameter $b_c$ from below.  The constant $\delta$ is the ratio of the critical impact parameter $b_c$ to the lengthscale $GM/v^2 = T_0 v/(2\pi)$.  Since $\eta \leq 1/4$ and we have assumed $\beta \ll 1$, we always have $\delta \ll 1$.  However, to satisfy our assumption of nonrelativistic orbits after the two black holes become bound, we need also to have $\beta_c \ll 1$, which is equivalent to the left-hand inequality of
\bb
\beta \ll \eta^{1/2} \leq 1/2.
\label{betarange}
\ee 
(The right-hand inequality, $\eta^{1/2} \leq 1/2$, follows immediately from the definition of the symmetric mass ratio $\eta \equiv M_1M_2/(M_1+M_2)^2$.)

Next, we can define dimensionless constants that depend not only on $\eta$ and $\beta$ but also on the dimensionless squared impact parameter $P = (b/b_c)^2$:
\bb
\beta_p \equiv P^{-1/2} \beta_c \equiv \left(\frac{96}{85\pi}\beta^2\frac{1}{\eta}\right)^{1/7} P^{-1/2} 
= \frac{2GM}{c^2\beta b_c P^{1/2}}
= \frac{2GM/(cv)}{b},
\label{betap}
\ee
\bb
\epsilon \equiv P^{-5/4} \delta 
\equiv \left(\frac{340\pi}{3}\beta^5\eta\right)^{1/7} P^{-5/4}
= \frac{\beta^2 b_c P^{-5/4}}{GM/c^2} = \frac{b P^{-7/4}}{GM/v^2}
= \left(\frac{85\pi}{24}\eta\right)^{1/2} \beta_p^{5/2}.
\label{epsilon}
\ee
The constant $\beta_p$ is, when it is small, a very good approximation for the dimensionless velocity maximum during the first bound orbit (i.e., at periapsis, the minimum separation distance $r_p$), namely $v_p/c$ for the initial bound orbit with impact parameter $b = b_c P^{1/2} \gg 4GM/(cv)$, with $4GM/(cv)$ being roughly the impact parameter below which there is direct coalescence without a period of nonrelativistic inspiral.  (The actual maximum impact parameter for direct coalescence is $4GM/(cv)$ multiplied by an order-one function of $\eta$ and $\beta$ that goes to 1 as both of those constants are taken to 0.)  By our general assumption that there are nonrelativistic inspiraling orbits after the two black holes become bound, we are requiring that $\beta_p \ll 1$, which is equivalent to $4GM/(cv) \ll b < b_c$, or
\bb
\frac{\beta^{4/7}}{\eta^{2/7}} \equiv \frac{(v/c)^{4/7}}{[M_1M_2/(M_1+M_2)^2]^{2/7}} \ll P \equiv \left(\frac{b}{b_c}\right)^2 < 1.
\label{Prange}
\ee
(Remember that $v$ is always the initial relative velocity of the two black holes when they are far apart and unbound, and $\beta = v/c$.)  Note that the nonrelativistic condition $\beta_p \ll 1$ implies that $\epsilon = (85\pi\eta/24)^{1/2}\beta_p^{5/2} \approx 1.667\,818(4\eta)^{1/2}\beta_p^{5/2} \ll 1$.
 
As we shall see shortly, the fractional change per orbit of the angular momentum given by Eq.\ (\ref{Deltal/l}) is proportional to $\epsilon^2$, as is the relative change per orbit of the eccentricity $e = \sqrt{1-f}$ and the absolute change of $f \equiv 1-e^2$ (though not the relative change of $f$, $E$, $a$, or $\tau$, which can be large per orbit; they are orbit parameters that do not change adiabatically when the orbit number $n$ is not large).  Therefore, for the angular momentum to change adiabatically, one needs $\epsilon \ll 1$, which is implied by the condition being assumed of nonrelativistic velocities at periapsis, $\beta_p \ll 1$, which is equivalent to $(\beta^2/\eta)^{2/7} \ll P = (b/b_c)^2$.

After defining these small constants characterizing the initial conditions, it is also convenient to define dimensionless measures of the time-dependent energy $E$ and angular momentum $L$, each of which are nearly constant for the major part of each orbit in which the separation distance $r$ between the black holes is much larger than the periapsis distance $r_p$ for that orbit, but which change mostly during the relatively short time when $r$ is near $r_p$.  That is, it is a good approximation to take the energy $E$ to have the constant value $E_n$ during most of the $n$th orbit and then rather suddenly change to $E_{n+1}$ as the next periapsis is passed, and similarly for the angular momentum $L$ to change rather suddenly from $L_n$ to $L_{n+1}$.

In particular, define the orbit-dependent (i.e., depending on the orbit number $n$) dimensionless variables
\bb
\lambda(n) \equiv \frac{L_n}{L_0} 
= \frac{L_n}{M\eta v b_c P^{1/2}},
\label{lambda}
\ee
which has the initial value $\lambda(0) = 1$, and
\bb
\nu(n) \equiv P^{7/2}\left(\frac{-E_n}{E_0}\right) 
= P^{7/2}\left(\frac{-2E_n}{M\eta v^2}\right).
\label{nu}
\ee
which has the initial value $\nu(0) = - P^{7/2}$.

In terms of the small constant $\epsilon^2$ and the variables $\nu(n)$ and $\lambda(n)$, the eccentricity factor for the $n$th orbit is
\bb
f(n) \equiv 1 - e(n)^2 = \epsilon^2 \lambda(n)^2 \nu(n).
\label{f(n)}
\ee

Although $\nu(n)$ has relatively large jumps as $n$ is increased, $\lambda(n)$ and $e(n)$ change relatively slowly (adiabatically), so Eq.\ (5.11) of \cite{Peters:1964zz} applies, which in my variables (suppressing the $n$ arguments for brevity henceforth for awhile) gives
\bb
\lambda = \left(1-f\right)^{3/19}
\left(1-\frac{121}{425}f\right)^{435/2299}
= 1 - \frac{18}{85}f + O(f^2).
\label{lambda-f}
\ee
Then one gets for the changes in $\lambda$ and $\nu$ for one orbit to be
\bb
\Delta\lambda = -\frac{18}{85} \epsilon^2 
\left(1-f\right)^{-12/19}
\left(1-\frac{121}{425}f\right)^{1740/2299}
\left(1-\frac{7}{15}f\right),
\label{Deltalambda}
\ee
\bb
\Delta\nu = \left(1-f\right)^{-21/19}
\left(1-\frac{121}{425}f\right)^{3045/2299}
\left(1-\frac{366}{425}f+\frac{31}{425}f^2\right)
= 1 + \frac{264}{425}f + O(f^2),
\label{Deltanu}
\ee
\bb
\Delta T = \tau_n = T_0 P^{21/4} \nu^{-3/2}\left[1 + O(\beta_p^2)\right].
\label{DeltaT}
\ee

For $\epsilon^2 \ll 1$, $\lambda$ stays near 1 for many orbits, but $\nu$ increases by very slightly more than 1 for each orbit, giving $f \approx \epsilon^2 (n - P^{7/2})$ for $n \ll 1/\epsilon^2 \sim \eta^{-2/7}\beta^{-5/7}P^{5/2}$.  Then Eq.\ (\ref{Deltanu}) implies that 
$\Delta\nu(n) \approx 1 + (264/425)\epsilon^2 (n - P^{7/2})$, which, along with the initial condition $\nu(0) = - P^{7/2}$, leads to
\bb
\nu(n) = - P^{7/2} + n + \frac{132}{425} \epsilon^2 (n^2 + n - 2nP^{7/2}) + O(\epsilon^4 n^3) + O(\beta_p^2 n).
\label{nuapprox}
\ee

Inserting this back into Eq.\ (\ref{DeltaT}) and summing over $n$ gives an estimate of the total inspiraling merger time as
\ba
T\!\! &\approx&\!\! \sum_{n=1}^\infty \tau_n = T_0 P^{21/4} \sum_{n=1}^\infty \nu(n)^{-3/2} 
\approx T_0 P^{21/4} \sum_{n=1}^\infty \left(-P^{7/4} + n + \frac{132}{425} \epsilon^2 n^2\right)^{-3/2}\nonumber \\
\!\!&\approx&\!\! \frac{2\pi GM}{v^3} P^{21/4} \left[\zeta\left(\frac{3}{2},1-P^{7/2}\right) - \frac{8}{5}\sqrt{\frac{33}{17}}
\left(\frac{340\pi}{3}\beta^5\eta\right)^{1/7} P^{-5/4} + O(\beta_p^2)\right]\!\!,
\label{Tformula}
\ea
using the Hurwitz zeta function,
\bb
\zeta(s,a) = \sum_{n=0}^\infty \frac{1}{(n+a)^s}.
\label{HZ}
\ee

If $\beta_p^2 \equiv [96/85\pi)]^{2/7}(\beta^2/\eta)^{2/7}P^{-1}$ and hence also $\epsilon^2 \equiv [(340\pi/3)\beta^5\eta]^{2/7}P^{-5/2}$ are small enough that we can neglect them, a highly precise simple elementary function approximation for the Hurwitz zeta function, using $z \equiv \zeta(3/2) \approx 2.612\,375$, gives
\bb
T \approx \frac{2\pi GM}{v^3}\left(\frac{P^{7/2}}{1-P^{7/2}}\right)^{3/2}\left[1 + (1-P^{7/2})^{3/2}\left(z-\frac{1-P^{7/2}}{1-0.5P^{7/2}}\right)\right],
\label{Ta}
\ee
which has a relative error always less than 0.07\% (assuming $\beta_p^2$ is negligible).

The appearance is a bit simpler in terms of
\bb
x \equiv 1 - P^{7/2} \equiv 1 - (b/b_c)^7
\label{x}
\ee
and a dimensionless time
\bb
t \equiv T/T_0 = t/(2\pi GM/v^3) \approx P^{21/4}\zeta(3/2,1-P^{7/2})
\label{t}
\ee
with the timescale $T_0 \equiv 2\pi GM/v^3$ from Eq.\ (\ref{T0}).  Then, again neglecting terms proportional to $\beta_p^2$ and $\epsilon^2$ and higher powers of these quantities, and using
\bb
z \equiv \zeta(3/2) \approx 2.612\,375,
\label{z}
\ee
one gets
\bb
T \equiv T_0\, t(P) \approx T_0 (1-x)^{3/2}\zeta(3/2,x) 
\approx \left(\frac{1-x}{x}\right)^{3/2}\left[1+x^{3/2}\left(z-\frac{2x}{1+x}\right)\right].
\label{Tax}
\ee


\section{Probability Distribution of the Inspiral Time}

From the inspiral merger time $T = T_0\, t(P)$ with $T_0 \equiv 2\pi G M/v^3$ when \\
$\beta_p^2 \equiv [96/85\pi)]^{2/7}(\beta^2/\eta)^{2/7}P^{-1}$ can be neglected, one can get the probability density per logarithmic interval $d\ln{T} = dT/T$ of the inspiral time $T$ as
\bb
\rho \equiv \frac{dP}{d\ln{T}} = t\frac{dP}{dt} \approx \frac{4}{21}P
\left[1 + \frac{P^{7/2}\zeta(5/2,1-P^{7/2})}{\zeta(3/2,1-P^{7/2})}\right]^{-1}.
\label{rho}
\ee

The following table gives numerical data, with the first column giving the cumulative conditional probability $P = (b/b_c)^2$ where $b_c$, given by Eq.\ (\ref{bc}), is the critical impact parameter for capture, the second column gives the dimensionless inspiral merger time $t = T/T_0 = T/(2\pi GM/v^3) \approx P^{21/4}\zeta(3/2,1-P^{7/2})$, the third column giving the inspiral time $T$ for $M_1 = M_2 = 5 M_\odot$ (so that $M = 10 M_\odot$ and $\eta = 1/4$) and for $v = 100$\ km/s $\approx (1/3)10^{-3}c$ (as well as estimates for $\beta_p^2 = (v_p/c)^2 \sim 2GM/r_p$ at periapsis), and the fourth column gives the probability density per logarithmic interval of the inspiral time $T$, namely $\rho = dP/d\ln{T} = dP/d\ln{t}$.  Note that with $\eta = 1/4$ and this value of $v$, one gets $\beta_p^2 \sim 0.01/P$, which is not very small when $P$ is small, giving uncalculated relative errors for $T$ of the order of $\beta_p^2$.  Thus the full 7 digits given for $t = T/T_0$ in the second column and for $\rho = dP/d\ln{T}$ in the fourth column are only all relevant for $\beta_p^2 \sim [\beta^2/(4\eta)]^{2/7}/P \stackrel{<}{\sim} 10^{-6}$ or $\beta \sim (4\eta)^{1/2}P^{7/4}\beta_p^{7/2} \stackrel{<}{\sim} 10^{-11}(4\eta)^{1/2} P^{7/4}$, but they are given for completeness. 

\vspace{3mm}


\begin{tabular}{||l|l|l|l||} \hline
$P=\left(\frac{b}{b_c}\right)^2$ & $t \equiv T/(2\pi GM/v^3)$ & $T$ for $10\,M_\odot$, 100 km/s 
& $\rho = \frac{dP}{d\ln{T}} = T\frac{dP}{dT}$ \\ \hline
0 & 0 & $\sim 2\pi GM/c^3 \sim 10^{-11}$ yr & 0 \\ \hline
0.01 &	$8.261\,057\times 10^{-11}$ & direct coalescence 
&0.001\,904\,762 \\ \hline
0.02 &	$3.143\,717\times 10^{-9}$ & direct coalescence 
& 0.003\,809\,522\\ \hline
0.05 & $3.860\,446\times 10^{-7}$ & nearly direct coalescence
& 0.009\,523\,673 \\ \hline
0.10 &	$1.469\,405\times 10^{-5}$ & $4\times 10^{-6}$ yr, $\beta_p^2 \sim 1/10$
& 0.019\,044\,525  \\ \hline
0.15 & $1.235\,812\times 10^{-4}$ & $3\times 10^{-5}$ yr, $\beta_p^2 \sim 1/15$
& 0.028\,552\,230  \\ \hline
0.20 & $5.605\,866\times 10^{-4}$ & $1.5\times 10^{-4}$ yr, $\beta_p^2 \sim 1/20$
& 0.038\,025\,045  \\ \hline
0.25 & $1.814\,879\times 10^{-3}$ &	$4.8\times 10^{-4}$ yr, $\beta_p^2 \sim 1/25$
& 0.047\,426\,785 \\ \hline
0.30 & $4.752\,460\times 10^{-3}$ &	$1.3\times 10^{-3}$ yr, $\beta_p^2 \sim 1/30$
& 0.056\,703\,625 \\ \hline
0.35 & $1.076\,525\times 10^{-2}$ & $2.8\times 10^{-3}$ yr, $\beta_p^2 \sim 1/35$
& 0.065\,780\,052 \\ \hline
0.40 & $2.196\,685\times 10^{-2}$ & $5.8\times 10^{-3}$ yr, $\beta_p^2 \sim 1/40$
& 0.074\,553\,181 \\ \hline
0.45 & $4.146\,907\times 10^{-2}$ & $1.10\times 10^{-2}$ yr, $\beta_p^2 \sim 1/45$
& 0.082\,884\,298 \\ \hline
0.50 & $7.380\,184\times 10^{-2}$ & $1.95\times 10^{-2}$ yr, $\beta_p^2 \sim 1/50$
& 0.090\,586\,038 \\ \hline
0.55 & $1.255\,964\times 10^{-1}$ & $3.32\times 10^{-2}$ yr, $\beta_p^2 \sim 1/55$
& 0.097\,402\,690 \\ \hline
0.60 & $2.067\,776\times 10^{-1}$ & $5.46\times 10^{-2}$ yr, $\beta_p^2 \sim 1/60$
& 0.102\,979\,918 \\ \hline
0.65 & $3.328\,098\times 10^{-1}$ &	$8.79\times 10^{-2}$ yr, $\beta_p^2 \sim 1/65$
& 0.106\,818\,463 \\ \hline
0.70 & $5.293\,246\times 10^{-1}$ & $1.40\times 10^{-1}$ yr, $\beta_p^2 \sim 1/70$
& 0.108\,205\,199 \\ \hline
0.75 & $8.427\,883\times 10^{-1}$ & $2.23\times 10^{-1}$ yr, $\beta_p^2 \sim 1/75$
& 0.106\,118\,525 \\ \hline
0.80 &	1.369\,021 & $3.62\times 10^{-1}$ yr, $\beta_p^2 \sim 1/80$
& 0.099\,129\,628 \\ \hline
0.85 & 2.347\,172 & $6.20\times 10^{-1}$ yr, $\beta_p^2 \sim 1/85$
& 0.085\,418\,810 \\ \hline
0.90 & 4.590\,029 &	1.21 years, $\beta_p^2 \sim 1/90$
& 0.063\,315\,302 \\ \hline
0.95 & $1.324\,718\times 10^1$ & 3.50 years, $\beta_p^2 \sim 1/95$
& 0.033\,208\,069 \\ \hline
0.98 & $5.265\,574\times 10^1$ & $1.39\times 10^1$ years, $\beta_p^2 \sim 0.01$
& 0.013\,250\,261 \\ \hline
0.99 & $1.500\,313\times 10^2$ & $3.96\times 10^1$ years, $\beta_p^2 \sim 0.01$
& 0.006\,619\,004 \\ \hline
0.999 &	$4.815\,761\times 10^3$ & $1.27\times 10^3$ years, $\beta_p^2 \sim 0.01$
& $6.655\,242\times 10^{-4}$ \\\hline
0.999\,9 & $1.526\,718\times 10^5$ & $4.03\times 10^4$ years, $\beta_p^2 \sim 0.01$
& $6.665\,281\times 10^{-5}$ \\ \hline
0.999\,99 &	$4.829\,293\times 10^6$ & $1.28\times 10^6$ years, $\beta_p^2 \sim 0.01$
& $6.666\,520\times 10^{-6}$ \\ \hline
0.999\,999 & $1.527\,202\times 10^8$ & $4.04\times 10^7$ years, $\beta_p^2 \sim 0.01$
& $6.666\,652\times 10^{-7}$ \\ \hline
0.999\,999\,9 & $4.829\,451\times 10^9$ & $1.28\times 10^9$ years, $\beta_p^2 \sim 0.01$
& $6.666\,665\times 10^{-8}$ \\ \hline
0.999\,999\,99 & $1.527\,207\times 10^{11}$ & $4.04\times 10^{10}$ years, $\beta_p^2 \sim 0.01$
& $6.666\,666\times 10^{-9}$ \\ \hline
0.999\,999\,999 &$ 4.829\,452\times 10^{12}$ & $1.28\times 10^{12}$ years, $\beta_p^2 \sim 0.01$
& $6.666\,667\times 10^{-10}$ \\ \hline
\end{tabular}

\newpage

\begin{figure}
\includegraphics[width=1.0\columnwidth]{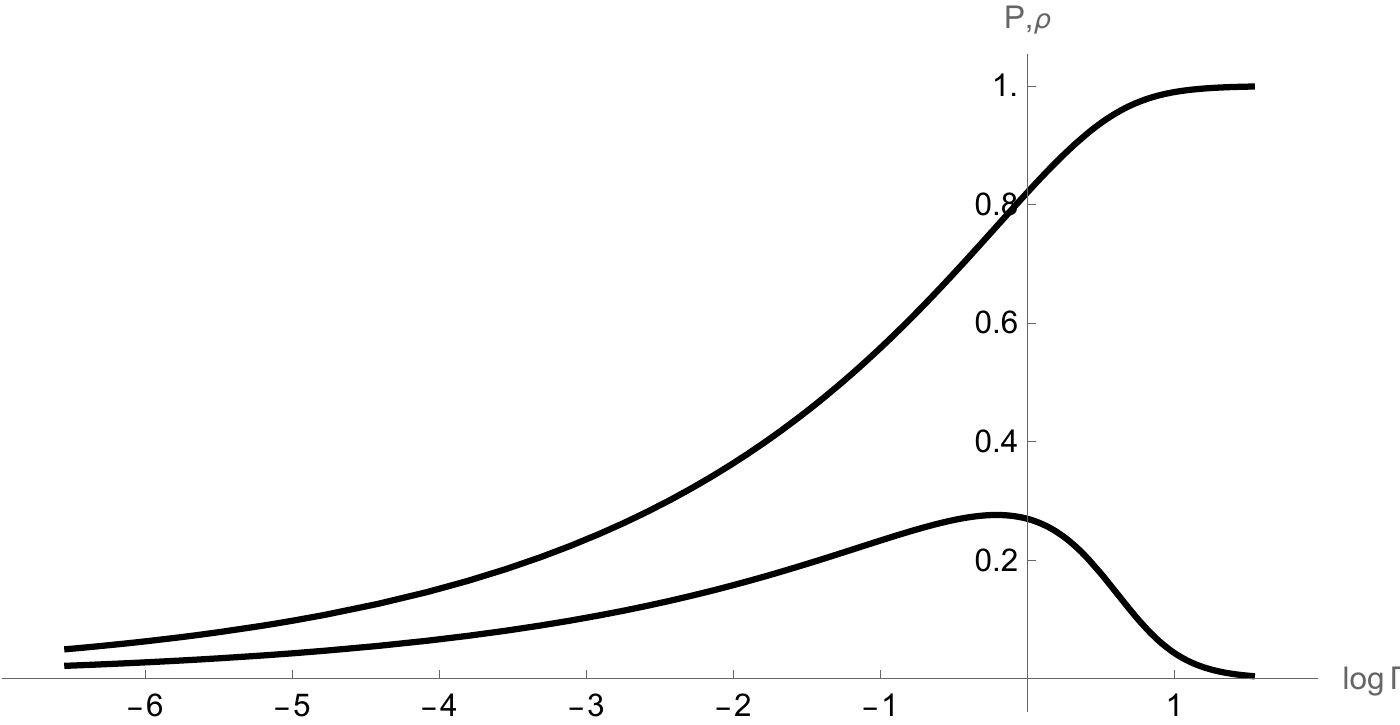}
\caption{The top graph gives the cumulative conditional probability $P = (b/b_c)^2$ for impact parameters smaller than $b$, given that they are smaller than the critical impact parameter $b_c = [340\pi G^7 M^7 \eta/(3 c^{14} \beta^9)]^{1/7}$ for capture into bound orbits, and the bottom graph gives $(\ln{10})*\rho = dP/d\log{t}$, both as functions of $\log{t}$, the common logarithm of the dimensionless inspiral time $t=T/(2\pi GM/v^3)$ with $T$ the inspiral time for two black holes of total mass $M$, symmetric mass ratio $\eta = M_1M_2/(M_1+M_2)^2$, and dimensionless incident relative velocity $\beta = v/c$.}
\end{figure}

As a function of the rescaled time 
$t=T/(2\pi GM/v^3) \approx P^{21/4}\zeta(3/2,1-P^{7/2})$,
the probability distribution tabulated in the Table and shown in Figure 1 is very broad, with $t$ varying by a factor of over a trillion ($1.816\times 10^{12}$) between the 1st and 99th percentiles (between $P=0.01$ and $P=0.99$), by a factor over 16 billion ($1.675\times 10^{10}$) between the 2nd and 98th percentile, by a factor over 34 million ($3.432\times 10^7$) between the 5th and the 95 percentile, by a factor over 312 thousand ($3.127\times 10^5$) between the 10th and 90th percentile, and by a factor slightly over 464 between the 25th and 75 percentiles (the first and third quantiles, $P=0.25$ and $P=0.75$).

The tails of the probability distribution per range of $\ln{t}$, $\rho = t dP/dt$, also decay slowly, asymptotically having the following form for very small and very large $t$ respectively, with $z \equiv \zeta(3/2) \approx 2.612\,375$:

\bb
\rho(t) = \frac{4}{21}\left(\frac{t}{z}\right)^{4/21}
\left[1-\frac{9}{7}\left(\frac{t}{z}\right)^{2/3}+O(t)\right],
\label{rhosmallt}
\ee
\bb
\rho(t) = \frac{4}{21}t^{-2/3}
\left[1-\frac{9}{7}t^{-2/3}+O(t^{-4/3})\right].
\label{rholarget}
\ee
These tails imply that the mean value of $t^s$ converges only for $-4/21 < s < 2/3$, though the first (left-hand) inequality in (\ref{Prange}), $P \gg (\beta^2/\eta)^{2/7}$, implies that the divergence at infinitesimal $t = T/T_0$ when $s < -4/21$ is not physical.  If one defines the inspiral time $T = T_0\,t = (2\pi GM/v^3)t$ so that it is always greater than $2GM/c^3$, the light travel time for a distance equal to the Schwarzschild radius corresponding to the total mass $M$, this would imply that one always has $\pi t > \beta^3$, which can be very small, but not infinitesimal, thus cutting off the divergence arising from the approximation $t = P^{21/4}\zeta(3/2,1-P^{7/2})$ that only applies for $t \gg (v/c)^3$.  

A quantity both of whose mean and standard deviation is finite is the logarithm of $t$, which has a mean value of $-3.899\,276$ and a standard deviation of $\ln{t}$ of 5.636\,468.  The exponential of the mean of $\ln{t}$ is 0.020\,0247, which is small, even smaller than the median value of $t$ of 0.073\,802 (at $P = 0.5$), but the exponential of the standard deviation of $\ln{t}$ is 280.470, another consequence of the breadth of the probability distribution for $t$ and $\ln{t}$.

Although Eq.\ (\ref{Ta}) gives an excellent closed-form approximation for $t(P)$ when $\epsilon$ is negligible (using $z \equiv \zeta(3/2) \approx 2.612\,375$),
\bb
t(P) \approx T \approx \left(\frac{P^{7/2}}{1-P^{7/2}}\right)^{3/2}\left[1 + (1-P^{7/2})^{3/2}\left(z-\frac{1-P^{7/2}}{1-0.5P^{7/2}}\right)\right],
\label{ta}
\ee
it is harder to find a good closed-form approximation for the inverse function $P(t)$, but one that always has relative error less than 1\% (again, neglecting $\beta_p^2$) is
\bb
P(t) \approx \left[1+\left(\frac{z^2 + (z-1)t}{zt + (z-1)t^2}\right)^{2/3}\right]^{-2/7}.
\label{Pb}
\ee

\section{Adiabatic Approximation for the Merger Time}

Now I shall give the adiabatic approximation for the merger time, $T_a$, as given by Eq.\ (\ref{Peters-eccentric}) of \cite{Peters:1964zz} from the values of the angular momentum $L_1 \approx \mu b v = \mu b_c P^{1/2} v$ and energy $E_1 \approx (1/2)\mu v^2 (1-P^{-7/2})$ during the first orbit, after the loss of energy and angular momentum to gravitational radiation during the first close encounter:

\ba
\mathrm{adiabatic\ merger\ time\ } \equiv T_a = T_0 t_a(P)
&\approx& \left(\frac{3}{85}\right)\frac{c^5 a_1^4 (1-e_1^2)^{7/2}}{G^3M^{3}\eta}\nonumber \\
&=& \left(\frac{3}{85}\right)\frac{c^5 L_1^7}{G^6M^{25/2}\eta^{15/2}(-2E_1)^{1/2}}\nonumber \\
&=& \left(\frac{3}{85}\right)\frac{c^5 v^6 P^{7/2} b_c^7}{G^6M^6\eta(P^{-7/2}-1)^{1/2}}\nonumber \\
&=& \frac{4\pi GM}{v^3}\frac{P^{7/2}}{(P^{-7/2}-1)^{1/2}}\nonumber \\
&=& 2T_0 P^{21/4}(1-P^{7/2})^{-1/2}\nonumber \\
&=& 2T_0(1-x)^{3/2}x^{-1/2}\nonumber \\
&=& 2x^{-1/2}[\zeta(3/2,x)]^{-1} T\nonumber \\
&\approx& 2x\left[1+x^{3/2}\left(z-\frac{2x}{1+x}\right)\right]^{-1} T,
\label{Tad}
\ea
where $T_0 \equiv 2\pi GM/v^3$, $P \equiv (b/b_c)^2$, $x \equiv 1-P^{7/2}$, $z \equiv \zeta(3/2)$, and the actual inspiral time is $T \approx T_0 P^{21/4} \zeta(3/2,1-P^{3/2}) = T_0 (1-x)^{3/2} \zeta(3/2,x)$ for $(\beta^2/\eta)^{2/7} \ll P < 1$.

The ratio $R_a$ of the adiabatic approximation for the inspiral merger time $T_a$ to the actual inspiral merger time $T$ (for $(\beta^2/\eta)^{2/7} \ll P < 1$ and neglecting third-body interactions) is
\bb
R_a \equiv \frac{T_a}{T} \approx 2x^{-1/2}[\zeta(3/2,x)]^{-1}
\approx 2x\left[1+x^{3/2}\left(z-\frac{2x}{1+x}\right)\right]^{-1}.
\label{Ra}
\ee
One can see that for $P=0$ or $x=1$, $R_a(0) = 2/\zeta(3/2) \approx 0.765\,587$, and as $P$ is increased to 1, $R_a(P)$ decreases monotonically to $R_a(0) = 0$.  Thus $R_a$ has a relative error a bit more than 23\% at $P=0$ but increasing to 100\% at $P=1$.

\newpage

\begin{figure}
\includegraphics[width=1.0\columnwidth]{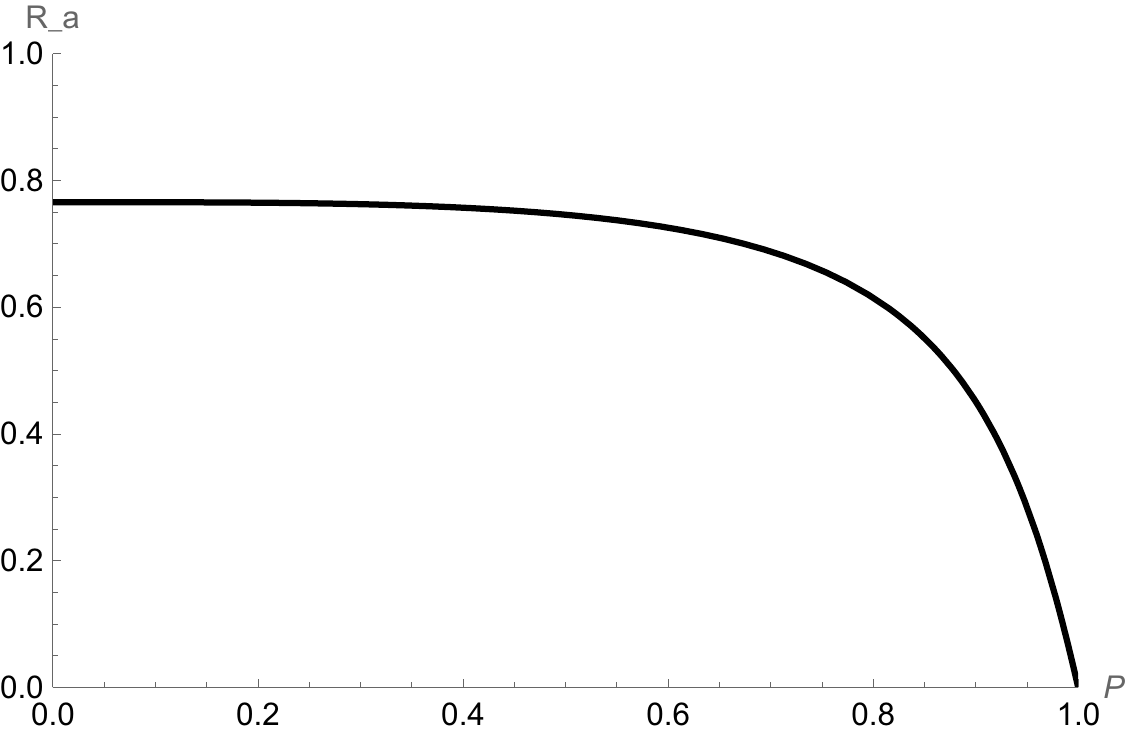}
\caption{This graph shows, as a function of the cumulative conditional probability $P = (b/b_c)^2$, the ratio $R_a = T_a/T$ of the adiabatic approximation $T_a$ for the inspiral merger time to the actual time $T$ for the inspiral merger time, using the discreteness of the orbits and the large relative changes in the periods of the first several orbits, which dominate the total inspiral time.  The adiabatic approximation is about 23\% too small for small $P$ but then becomes 100\% too small as $P$ is increased to 1.}
\end{figure}

\begin{figure}
\includegraphics[width=1.0\columnwidth]{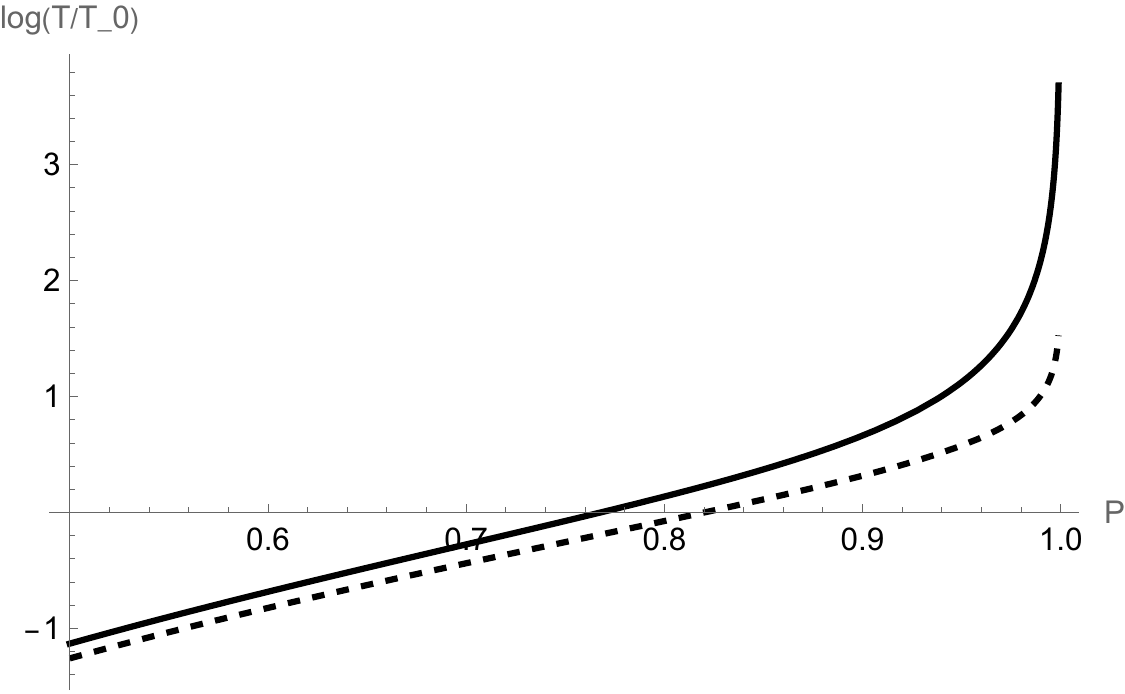}
\caption{The solid curve gives the common logarithm (base 10) of the dimensionless time $t = T/T_0 \approx P^{21/4}\zeta(3/2,1-P^{7/2})$ for the inspiral time as the sum of the times of the discrete orbits as a function of the cumulative conditional probability $P = (b/b_c)^2$, whereas the dashed curve gives the common logarithm of the adiabatic approximation $t_a = T_a/T_0 \approx 2P^{21/4}(1-P^{7/2})^{-1/2}$ for the inspiral time using the adiabatic continuum approximation of Peters.}
\end{figure}

\begin{figure}
\includegraphics[width=1.0\columnwidth]{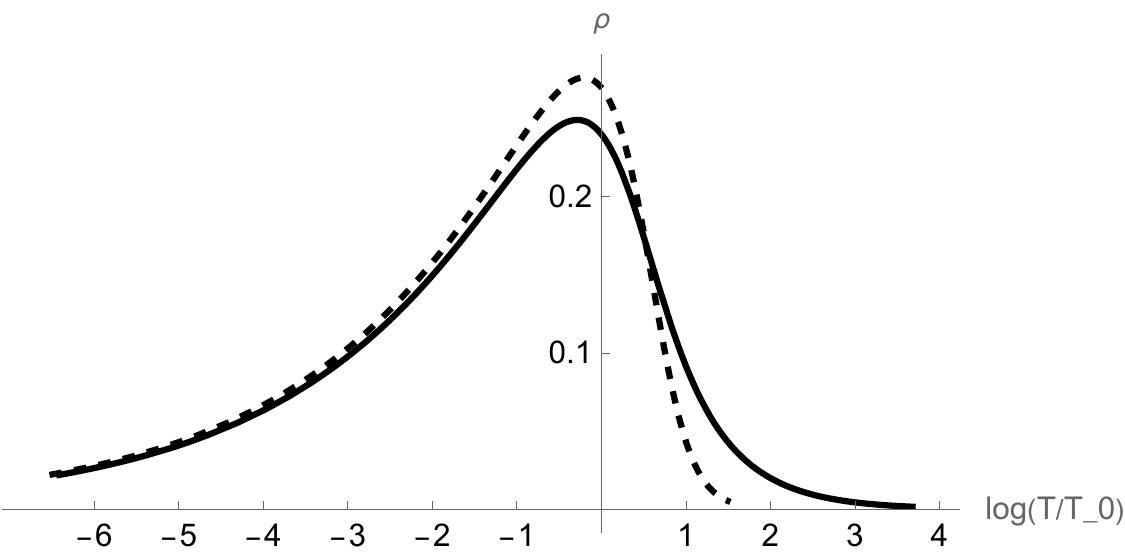}
\caption{The solid curve, the same as the lower curve of Figure 1, though here with an expanded vertical scale, gives $(\ln{10})*\rho = dP/d\log{t}$ as a function of $\log{t}$, the common logarithm of the dimensionless inspiral time $t=T/(2\pi GM/v^3)$ with $T$ the actual predicted inspiral time for two black holes of total mass $M$, symmetric mass ratio $\eta = M_1M_2/(M_1+M_2)^2$, and dimensionless incident relative velocity $\beta = v/c$.  The dashed curve gives
$(\ln{10})*\rho_a = dP/d\log{t_a}$ as a function of $\log{t_a}$, the common logarithm of the dimensionless inspiral time 
$t_a=T_a/(2\pi GM/v^3)$ with $T_a$ the inspiral time as predicted by the invalid adiabatic approximation.  Note that points on the two graphs at the same horizontal value of $\log{t} = \log{t_a}$ do not have the same value of $P$ and hence do not have the same impact parameter $b = b_c \sqrt{P}$.}
\end{figure}

\newpage 

Unlike the inspiral merger time $T(P)$ from the set of discrete orbits in which the orbit time $\tau_n$ jumps significantly as $n$ increases by 1 from one orbit to the next (until $n$ gets large, since the fractional change in $\tau_n$ is approximately $-3/(2n)$ for large $n$), which has a divergent mean, the adiabatic approximation giving $T_a(P)$ for the merger time incorrectly gives a finite mean value for it:
\bb
\langle T_a \rangle \approx \frac{11\sqrt{\pi}}{9}\frac{\Gamma(11/14)}{\Gamma(2/7)}\,T_0 \approx 0.812\,204\, \frac{2\pi GM}{v^3}.
\label{Tamean}
\ee
However, the standard deviation of the adiabatic approximation $T_a$ does diverge as it also does for the correct $T$.  Whereas the mean value of $T^s$ converges only for $-4/21 < s < 2/3$, the mean value of $T_a^s$ converges for $-4/21 < s < 2$, a broader range, but not quite broad enough to include $s=2$ as would be necessary for the standard deviation to converge.

Both $T$ and $T_a$ have finite mean values for powers of their logarithms, so they give one way to get finite comparisons of the adiabatic approximation $T_a$ with the actual $T$ from the discrete orbits.  For example, subtracting out $\ln{T_0}$, whereas $\ln{t}$ has a mean value of $-3.899\,276$, whose exponential is 0.020\,025, $\ln{t_a}$ has a mean value of $-4.361\,004$, whose exponential is 0.012\,766, which is a bit less than 64\% as large as that for $\ln{t}$.  Furthermore, whereas $\ln{t}$ has a standard deviation of 5.636\,468, whose exponential is 280.470, $\ln{t_a}$ has a standard deviation of 5.433\,399, whose exponential is 226.648, a bit less than 81\% as large as that of $\ln{t}$.  Also, the median value of $t_a$, its value for $P=1/2$, is 0.055\,045, which is somewhat less than 75\% as large as that of $t$.  Thus the adiabatic approximation is not a very good approximation for the inspiral merger time, especially for impact parameters $b$ near the critical impact parameter $b_c$, where $P = (b/b_c)^2$ is near 1 and $R_a = t_a/t \sim 2(1-P^{7/2}) = 2[1-(b/b_c)^7]$ approaches 0, so that the actual inspiral time becomes arbitrarily larger than what the adiabatic approximation gives.

The main reason that the adiabatic approximation gives an inspiral merger time $T_a = T_0 t_a$ significantly less than the more accurate calculation giving an inspiral merger time $T = T_0 t$ from discrete orbits is that the period $\tau$ of each orbit is proportional to the 3/2 power of the semimajor axis $a$, and this changes almost entirely only near the periapsis at the beginning of each orbit, where most of the gravitational radiation is emitted.  However, the adiabatic approximation assumes that the semimajor axis is decreasing at a smooth monotonic rate, decreasing the period below what actually occurs with the semimajor axis decreasing mainly just near the beginning of each orbit.

In particular, when one can neglect $\beta_p^2 \equiv [96/85\pi)]^{2/7}(\beta^2/\eta)^{2/7}P^{-1} = 2GM/(cvb)$, one can write $T$ and $T_a$ in similar forms for comparison (again with $T_0 \equiv 2\pi GM/v^3$ and $P \equiv (b/b_c)^2$):
\bb
T \approx T_0 P^{21/4}\sum_{n=1}^\infty (n-P^{7/2})^{-3/2} = T_0 P^{21/4} \zeta(3/2,1-P^{7/2}),
\label{Tsum}
\ee
\bb
T_a \approx T_0 P^{21/4}\int_1^\infty d\nu (\nu-P^{7/2})^{-3/2} = 2 T_0 P^{21/4} (1-P^{7/2})^{-1/2}.
\label{Taintegral}
\ee
Thus the fact that the adiabatic approximation $T_a$ is significantly smaller than the valid calculation with discrete orbits can be traced to the fact that in the part of the integral from $\nu = n$ to $\nu = n+1$, the average value of $(\nu-P^{7/2})^{-3/2}$ is less than the $(n-P^{7/2})^{-3/2}$ it would need to be for each $n$ for the adiabatic integral for $T_a$ to equal the discrete orbit sum for $T$.  The difference,
\bb
T-T_a \approx T_0 P^{21/4}\sum_{n=1}^\infty \int_n^{n+1} d\nu[(n-P^{7/2})^{-3/2}-(\nu-P^{7/2})^{-3/2}],
\label{T-Ta}
\ee
is analogous to the Euler constant
\bb
\gamma = \sum_{n=1}^\infty \int_n^{n+1} d\nu[n^{-1}-\nu^{-1}].
\label{gamma}
\ee

The sum in Eq.\ (\ref{T-Ta}) for $T-T_a$ is dominated by the $n=1$ term, which corresponds to the first bound orbit.  This difference, and also the ratio $T/T_a$, diverges in the limit $P \equiv (b/b_c)^2 \rightarrow 1$, which corresponds to the impact parameter $b$ approaching the critical impact parameter $b_c$ from below, leading to a divergence in the semimajor axis $a_1$ and in the period $\tau_1 = 2\pi\sqrt{a_1^3/(GM)}$ of the first orbit in this limit.

\section{Conclusions}

Although the formulas of Peters and of Peters and Mathews \cite{Peters:1963ux,Peters:1964qza,Peters:1964zz} work well for calculating the critical impact parameter $b_c \approx [340\pi \eta/(3 \beta^9)]^{1/7}(GM/c^2)$ \cite{Quinlan:1989,Mouri2002,OLeary:2008myb,Page:2024zpr} for two black holes of total mass $M = M_1+M_2$, symmetric mass ratio $\eta = M_1M_2/(M_1+M_2)^2$, nonrelativistic initial relative velocity $v = \beta c \ll \eta^{1/2} c$, impact parameter $b = P^{1/2}b_c$ and nonrelativistic periapsis velocity $v_p = \beta_p c \approx 2GM/(bv)\ll c$, the adiabatic approximation of \cite{Peters:1964zz} (assuming that the fractional changes of the semimajor axis and period are small from one orbit to the next) that has been widely used
\cite{Lee1993,OLeary:2008myb,Clesse:2016ajp,Kovetz:2016kpi,Ali-Haimoud:2017rtz,Gondan:2017wzd,Sasaki:2018dmp,Rodriguez:2018pss,Raidal:2018bbj,Vaskonen:2019jpv,Korol:2019jud,Jedamzik:2020ypm,Kritos:2020fjw,Tucker:2021mvo,Kocsis2022}
significantly underestimates the inspiral merger time, with relative errors ranging from approximately $1-2/\zeta(3/2)$ (a bit over 23\%) for $b \ll b_c$ to 100\% in the limit as $b$ approaches $b_c$.  However, for $2GM/(cv) \ll b \leq b_c$, most of the inspiral time is spent in nonrelativistic Keplerian orbits with eccentricity $e$ near 1, so that most of the gravitational radiation occurs near periapsis, and one can approximate the dominant sequence of orbits as having the angular momentum and minimum (periapsis) separation distance $r_p$ stay nearly constant for this sequence of orbits, but the nonrelativistic energy $E = -GM^2\eta/(2a)$ decreasing by a nearly constant amount $-\Delta E \approx (1/2)\mu v^2 P^{-7/2} = (170\pi/3)[GM/(cvb)]^7\eta^2Mc^2$ from one orbit to the next, and hence the semimajor axis $a = GM^2\eta/(-2E)$ and corresponding period $\tau = 2\pi\sqrt{a^3/(GM)}$ decreasing by calculable amounts from one orbit to the next.  In this way one can use the formulas of Peters and of Peters and Mathews \cite{Peters:1963ux,Peters:1964qza,Peters:1964zz} to calculate the total inspiral merger time as $T \approx (2\pi GM/v^3)P^{21/4}\zeta(3/2,1-P^{7/2})$, where $P = (b/b_c)^2$.

\section*{Acknowledgments}

This work was motivated by research by Ashaduzzaman Joy on the number of gravitons emitted by black-hole coalescences observed by LIGO and Virgo.  Helpful suggestions by the editor pointed out \cite{Hansen:1972jt} and other relevant papers and have led to many significant improvements in the manuscript.  Clifford Will kindly alerted me to \cite{Walker:1979xk} and \cite{Tucker:2021mvo}.  Financial support was provided by the Natural Sciences and Engineering Research Council of Canada.

\end{document}